\documentclass[twocolumn,showpacs,superscriptaddress,prl,aps,floatfix]{revtex4-1}
\usepackage{graphicx}
\usepackage{rotating}
\usepackage{amsmath}
\usepackage[usenames,dvipsnames]{color}
\usepackage{float}
\usepackage{rotating}
\usepackage{booktabs}
\usepackage{multirow}
\usepackage{comment}
\usepackage[colorlinks,linkcolor=magenta,citecolor=blue]{hyperref}
\usepackage[T1]{fontenc}
\usepackage{bm}
\usepackage{setspace}
\usepackage{xr-hyper}
\makeatletter

\newcommand{\Rmnum}[1]{\expandafter\@slowromancap\romannumeral #1@}

\hfuzz=\maxdimen
\begin{document}
	
	  \title{Polarization Rotation Drives a Spin-Topological Transition in Ferroelectric Bismuth Monolayer}
	
	\author{Jinming Zhai}
	\affiliation{Key Laboratory of Advanced Materials and Devices for Post-Moore Chips, Ministry of Education, Beijing Key Laboratory for Magneto-Photoelectrical Composite and Interface Science, School of Mathematics and Physics, University of Science and Technology Beijing, Beijing 100083, China}
	
	\author{Lingzhi Cao}
    \affiliation{Key Laboratory of Advanced Materials and Devices for Post-Moore Chips, Ministry of Education, Beijing Key Laboratory for Magneto-Photoelectrical Composite and Interface Science, School of Mathematics and Physics, University of Science and Technology Beijing, Beijing 100083, China}

	\author{Yateng Wang}
    \affiliation{Key Laboratory of Advanced Materials and Devices for Post-Moore Chips, Ministry of Education, Beijing Key Laboratory for Magneto-Photoelectrical Composite and Interface Science, School of Mathematics and Physics, University of Science and Technology Beijing, Beijing 100083, China}	
	
	\author{Huicong Li}
	\affiliation{Key Laboratory of Advanced Materials and Devices for Post-Moore Chips, Ministry of Education, Beijing Key Laboratory for Magneto-Photoelectrical Composite and Interface Science, School of Mathematics and Physics, University of Science and Technology Beijing, Beijing 100083, China}
	
	\author{Zhilong Yang}
	\affiliation{Key Laboratory of Advanced Materials and Devices for Post-Moore Chips, Ministry of Education, Beijing Key Laboratory for Magneto-Photoelectrical Composite and Interface Science, School of Mathematics and Physics, University of Science and Technology Beijing, Beijing 100083, China}

	\author{Yali Yang}
    \email{ylyang@ustb.edu.cn}
    \affiliation{Key Laboratory of Advanced Materials and Devices for Post-Moore Chips, Ministry of Education, Beijing Key Laboratory for Magneto-Photoelectrical Composite and Interface Science, School of Mathematics and Physics, University of Science and Technology Beijing, Beijing 100083, China}
	
	\author{Jiangang He}
	\email{jghe2021@ustb.edu.cn}
	\affiliation{Key Laboratory of Advanced Materials and Devices for Post-Moore Chips, Ministry of Education, Beijing Key Laboratory for Magneto-Photoelectrical Composite and Interface Science, School of Mathematics and Physics, University of Science and Technology Beijing, Beijing 100083, China}
	
	\date{\today}
	
	\begin{abstract}
    Bismuth monolayer is the first two-dimensional elemental ferroelectric and an appealing platform for coupling polar order to spin-orbit-driven topology. However, its microscopic switching mechanism remains elusive. Here, using first-principles lattice dynamics and symmetry-adapted mode analysis, we identify a previously overlooked rotational pathway for in-plane polarization switching. Its energy barrier is more than four times lower than that of direct reversal, naturally explaining the vortexlike domain textures observed in molecular dynamics simulations. Remarkably, this polarization rotation also drives a spin-topological transition, changing the spin Chern number from $C_s=-2$ to $0$. Directional uniaxial strain further steers the polarization orientation and tunes the associated topological transition. These results establish polarization rotation as the switching mechanism of ferroelectric Bi monolayer and as an efficient route to electrically and mechanically programmable topology in two-dimensional ferroelectrics.
	\end{abstract}

	\maketitle
	
	Ferroelectric materials, characterized by a spontaneous electric polarization that can be reversed by an external electric field, have long attracted intense interest because of their fundamental significance~\cite{rabe2007modern,strukov2012ferroelectric}, strong coupling between polar distortions and other order-parameters~\cite{https://doi.org/10.1002/adma.201203199,he2018tunable,doi:10.1021/jacs.4c03296}, and broad technological applications~\cite{doi:10.1126/science.1129564,martin2016thin,RevModPhys.77.1083}. The recent realization of robust ferroelectricity in atomically thin crystals has pushed this physics to the two-dimensional (2D) limit~\cite{doi:10.1126/science.aad8609,Wang2023NatMater,yuan2019room,fei2018ferroelectric}. Beyond device scaling and energy-efficient nanoelectronics, reduced dimensionality enhances the coupling among polarization, lattice distortions, electronic structure, and interfaces, creating opportunities for ferroelectric phenomena with no direct bulk analogue~\cite{Wang2023NatMater,Zhang2023NatRevMater,RevModPhys.93.011001,https://doi.org/10.1002/aelm.201900818}.
	
	Elemental group-VA monolayers provide a particularly unconventional route to 2D ferroelectricity. Unlike conventional ferroelectrics, where polar order typically originates from relative displacements between chemically distinct cation and anion sublattices, puckered As, Sb, and Bi monolayers were predicted to become polar through spontaneous lattice buckling~\cite{Xiao2018AFM}. Bismuth monolayer is the experimentally realized representative of this family: its black-phosphorus-type ($Pmn2_1$) phase hosts in-plane ferroelectricity arising from charge redistribution, symmetry-lowering lattice distortion, and active lone-pair electrons~\cite{Gou2023Nature}. This elemental ferroelectric further exhibits strain-tunable charged domain walls~\cite{Zhong2024NatCommun}, giant negative in-plane piezoelectricity, and strong nonlinear optical responses~\cite{Zhong2023PRL,Wang2023PRB}. Recent molecular-dynamics (MD) simulations have revealed rich phase evolution, domain dynamics, thermal transport anomalies, and vortexlike topological textures in Bi monolayers~\cite{Zhang2024PRL,Hong2025PRL}, while related dynamical approaches are being developed for 2D IV--VI ferroelectrics~\cite{Yu2026PRB}.
	
	A second key feature that makes monolayer Bi especially compelling is its strong spin-orbit coupling (SOC). Bi-based 2D systems have long served as paradigmatic platforms for quantum spin Hall physics~\cite{Murakami2006PRL,Drozdov2014NatPhys,Reis2017Science}. In ferroelectric Bi monolayers, theory predicts enhanced Berry-curvature dipoles (BCD) and persistent spin textures~\cite{Jin2021NanoLett}. Photoexcitation can drive transient phases with distinct electronic and spin-topological characters~\cite{Peng2024PRL}. Bismuth bilayers exhibit sliding-switchable topology, geometric currents, and BCD responses~\cite{Qian2025CPL}, and elemental group-VA 2D ferroelectrics can generate sizable shift-current and second-harmonic responses~\cite{YoungRappe2012PRL,Qian2023NPJ,Chen2026SHG}. These developments suggest that monolayer Bi may provide a rare elemental platform in which ferroelectric order, Berry geometry, nonlinear response, and spin topology can be controlled by the same structural degrees of freedom.
	
	Despite these advances, the microscopic mechanism of polarization switching in monolayer Bi remains elusive. The conventional picture of ferroelectric switching often assumes a one-dimensional double-well potential, in which polarization reversal proceeds through a high-symmetry paraelectric phase (e.g., $Pmna$ in Bi monolayer). Whether this picture applies to elemental Bi monolayer is unclear. In particular, the experimentally observed ferroelectric phase is $Pmn2_1$~\cite{Gou2023Nature}, whereas several competing low-energy structures have been identified computationally~\cite{Singh2019JPCL}. This raises a fundamental question: does polarization reversal in monolayer Bi occur through a direct high-symmetry path, or through a multidimensional low-symmetry pathway with a much lower energy barrier? Resolving this question is essential not only for understanding ferroelectric switching itself, but also for explaining the vortexlike domain textures observed in large-scale MD simulations~\cite{Hong2025PRL} and for determining whether structural switching can be used to control spin topology.

	\begin{figure*}[!htb]
	\includegraphics[clip,width=1.0\linewidth]{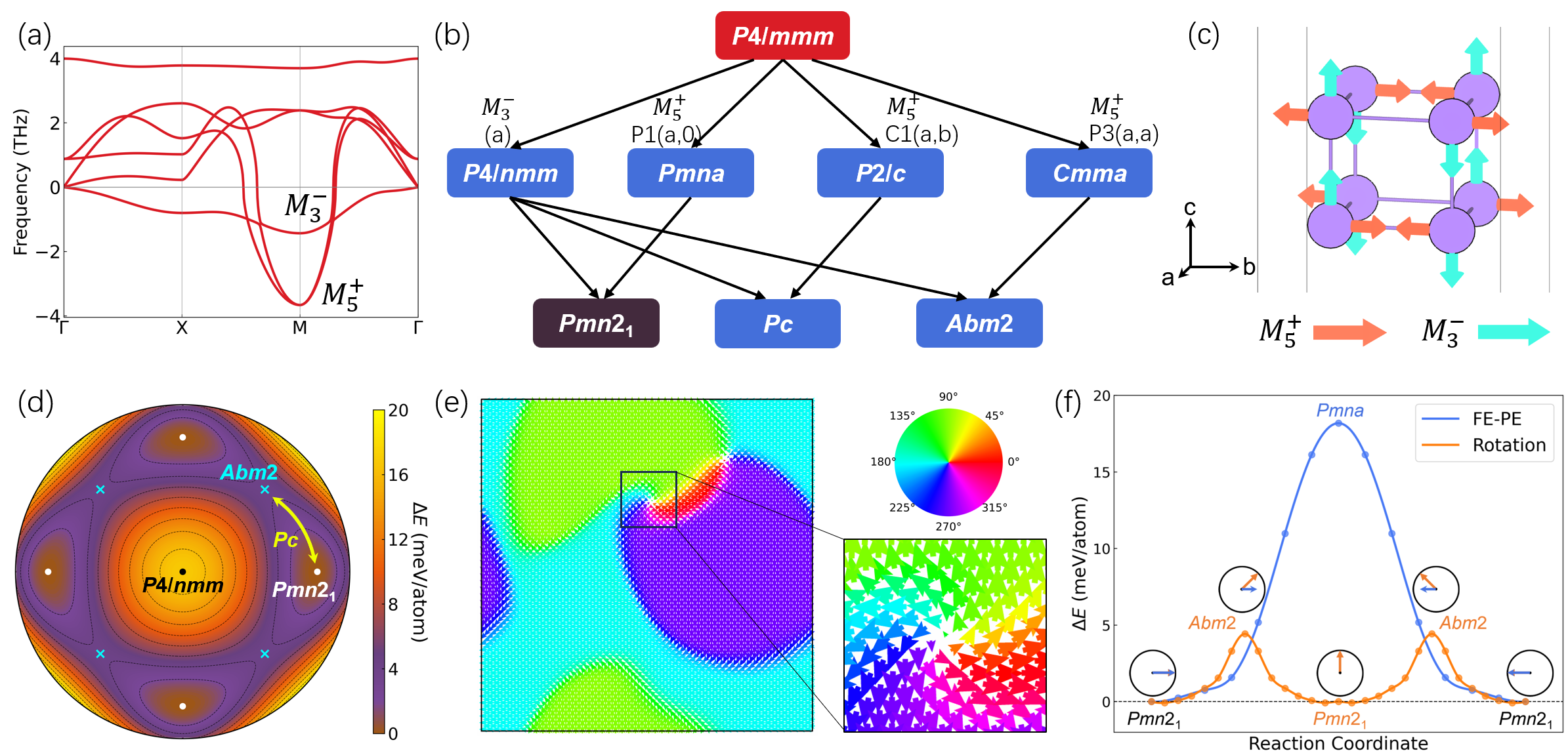}
	\caption{Soft-mode origin of the multidimensional energy landscape in Bi monolayer. (a) Phonon spectrum of the $P4/mmm$ phase. (b) Group-subgroup tree starting from the $P4/mmm$ phase through the unstable $M_3^-$ and $M_5^+$ modes and their coupled descendants. (c) Schematic atomic displacement patterns associated with the $M_5^+$ and $M_3^-$ soft modes. (d) 2D energy contour in the $(Q_a,Q_b)$ space of the $M_5^+$ order parameter, with the energy zero referenced to the $Pmn2_1$ minimum. (e) Transient vortexlike structures during the transition from the $P4/nmm$ phase to the ferroelectric state. The colored arrows indicate the direction and relative magnitude of the local order-parameter. (f) Comparison between direct ferroelectric reversal and rotation-mediated switching in Bi monolayer. The inserted arrows schematically indicate the calculated in-plane polarization direction and relative magnitude for representative structures along the path. The initial $Pmn2_1$ state is taken as the zero-energy reference for both paths.}
	\label{fig1}
	\end{figure*}

	Here we show that polarization switching in Bi monolayer is governed not by a one-dimensional double well, but through polarization rotation in a multidimensional soft-mode energy landscape. Starting from the high-symmetry $P4/mmm$ parent phase, first-principles lattice-dynamical calculations and symmetry-adapted mode analysis identify the experimentally observed $Pmn2_1$ phase as the energy minimum, the $Abm2$ phase as a saddle point connecting two ferroelectric states with mutually perpendicular polarization directions, and the $Pc$ structure as the low-energy transition channel between them. This landscape reveals a rotational switching pathway with an ultralow barrier of $4.4$ meV/atom, far below the $18.2$ meV/atom barrier of the direct polarization reversal. The same rotational pathway naturally accounts for the vortexlike domain textures observed in MD simulations. More importantly, polarization rotation drives a spin-topological transition through band-gap closing and reopening, accompanied by a change of the spin Chern number from $-2$ to $0$. Finally, our large-scale MD simulations show that directional uniaxial strain selects the order-parameter orientation and induces phase transition between $Pmn2_1$ and $Abm2$. Bismuth monolayer therefore realizes an electrically and mechanically programmable 2D ferroelectric in which polarization rotation, topology, Berry geometry, and domain order are governed by a common structural order parameter.

	All first-principles calculations were carried out within density functional theory (DFT) using the Vienna \textit{Ab initio} Simulation Package (\textsc{VASP})~\cite{Kresse1993,Kresse1996,Kresse1996b}, and phonon spectra were obtained by the finite-displacement method as implemented in \textsc{Phonopy}~\cite{Togo2015}. The projector augmented-wave (PAW) method~\cite{PAW1,PAW2} was adopted together with the PBEsol exchange--correlation functional~\cite{pbesol1,Perdew2008PRL}. Minimum-energy switching pathways were determined using the climbing-image nudged elastic band (CI-NEB) method~\cite{Henkelman2000,Henkelman2000b}. Symmetry-adapted mode analysis and the construction of Landau-type invariants were performed with ISOTROPY~\cite{Campbell2006,Stokes_ISODISTORT,Stokes1988,Hatch2003INVARIANTS}. Topological and Berry-geometric properties, including Wilson loops, Berry curvature, and BCD, were evaluated from Wannier-based tight-binding Hamiltonians using Wannier90 and WannierTools~\cite{Mostofi2008,Mostofi2014,Wu2018}. Further details on the computational details, mode decomposition, Landau expansion, topological analysis, and validation of the deep-potential model are provided in the Supplemental Material~\cite{SM}.
	\nocite{Soluyanov2011PRB}

	To identify the structural origin of ferroelectricity in Bi monolayer, we start from the high-symmetry $P4/mmm$ parent phase and follow its lattice instabilities. As shown in Fig.~\ref{fig1}(a), the phonon spectrum contains three unstable branches at the $M$ point: a doubly degenerate $M_5^+$ mode and a nondegenerate $M_3^-$ mode. The $M_3^-$ instability consists of antipolar out-of-plane Bi displacements and lowers the symmetry to $P4/nmm$ [Fig.~\ref{fig1}(b)]. The $M_5^+$ mode, in contrast, is an in-plane antipolar displacement of Bi atoms and therefore forms a two-component structural order parameter. Condensing this mode along different order-parameter directions generates the $Pmna$, $P2/c$, and $Cmma$ subgroups, associated with the $P1(a,0)$, $C1(a,b)$, and $P3(a,a)$ directions, respectively. The corresponding displacement patterns of the $M_5^+$ and $M_3^-$ soft modes are illustrated in Fig.~\ref{fig1}(c).
	
	The relevant low-energy structures emerge when the in-plane $M_5^+$ distortion couples to the $M_3^-$ mode. This coupling further lowers the symmetry and produces the $Pmn2_1$, $Pc$, and $Abm2$ structures shown in Fig.~\ref{fig1}(b). The experimentally observed ferroelectric $Pmn2_1$ phase~\cite{Gou2023Nature} is thus naturally understood as a soft-mode-derived state. We construct a Landau free-energy model in terms of the two components of the $M_5^+$ order parameter, $(Q_a,Q_b)$, and map the associated energy surface. The mode amplitudes and explicit Landau expansion are given in Table~S1 and Eqs.~(S1)--(S5)~\cite{SM}. Because the essential order-parameter is two-dimensional, the switching physics cannot be reduced to a conventional one-dimensional double well. It is instead governed by a 2D energy landscape in the $(Q_a,Q_b)$ space spanned by the symmetry-equivalent components of the unstable $M_5^+$ mode.
	
	Figure~\ref{fig1}(d) shows the resulting clamped-lattice energy contour in the $(Q_a,Q_b)$ space, with the $M_3^-$ amplitude optimized at each point (see Sec.~II of the Supplemental Material~\cite{SM}). The four white dots mark symmetry-equivalent $Pmn2_1$ minima, corresponding to four ferroelectric states with distinct polarization orientations. The four red crosses denote $Abm2$ saddle points connecting neighboring $Pmn2_1$ variants whose polarization directions differ by $90^\circ$. The low-symmetry $Pc$ structures form the low-energy channels away from the high-symmetry directions, with arbitrary ratios of $Q_a/Q_b$ ($Q_a/Q_b$ $\neq$ 0, $\pm$1, $\infty$). Writing the order parameter as $(Q_a,Q_b)=Q(\cos\theta_Q,\sin\theta_Q)$, the four $Pmn2_1$ states occur at $\theta_Q=0^\circ$, $90^\circ$, $180^\circ$, and $270^\circ$, whereas the four $Abm2$ states occur at $\theta_Q=45^\circ$, $135^\circ$, $225^\circ$, and $315^\circ$. This multidimensional landscape rationalizes the multiple closely related low-energy phases previously identified by minima hopping~\cite{Singh2019JPCL} and is further supported by our large-scale MD simulations [Fig.~\ref{fig1}(e)]. During the structural evolution from the $P4/nmm$ reference phase to the ferroelectric state at $0.1$ K, the system explores several symmetry-related directions in order-parameter space, forming transient local-vortex textures before finally condensing into the lowest-energy $Pmn2_1$ domains. Such vortexlike textures, also observed in recent MD simulations of Bi monolayers~\cite{Hong2025PRL}, provide direct dynamical evidence for the soft-mode-driven multidimensional energy landscape established here. Additional size-dependent vortex maps are shown in Figs.~S6 and S7~\cite{SM}. This energy landscape immediately suggests that polarization switching can proceed by rotation rather than by collinear reversal.
	
	We further confirm this mechanism by comparing two CI-NEB pathways, as shown in Fig.~\ref{fig1}(f). The direct ferroelectric--paraelectric--ferroelectric path, connecting two oppositely polarized $Pmn2_1$ states through the paraelectric $Pmna$ structure, has an energy barrier of $18.2$~meV/atom. By contrast, the rotation-mediated path, in which the polarization in the $Pmn2_1$ phase rotates by $90^\circ$ through the $Abm2$ saddle, has a barrier of only $4.4$~meV/atom. Polarization rotation in the 2D order-parameter space is therefore energetically preferred over direct collinear reversal, establishing a low-barrier switching mechanism for ferroelectric Bi monolayer.

	\begin{figure}[tph!]
		\includegraphics[clip,width=1.0\linewidth]{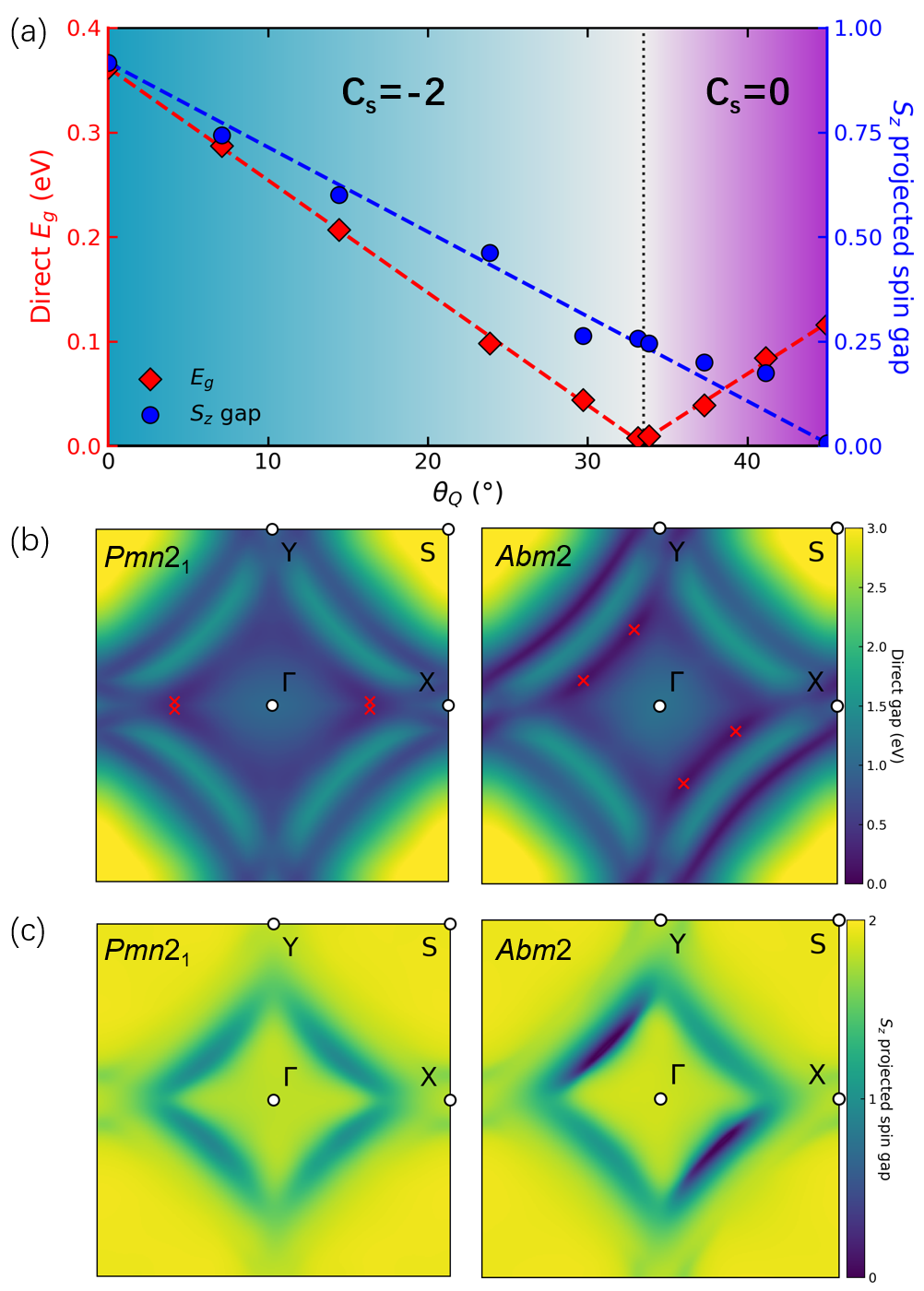}
		\caption{Topological transition and gap reconstruction along the rotational switching path.
			(a) Evolution of the direct band gap ($E_{g}$), projected $S_z$ spin gap, and spin Chern number $C_s$ along the rotational pathway. (b) $\mathbf{k}$-resolved direct band gap over the first Brillouin zone for $Pmn2_1$ and $Abm2$. The red crosses mark the momentum positions where the direct band gap reaches its minimum for each structure. (c) $\mathbf{k}$-resolved projected $S_z$ spin gap over the first Brillouin zone for $Pmn2_1$ and $Abm2$.}
		\label{fig2}
	\end{figure}

	More importantly, the low-barrier polarization-rotation pathway is not only a structural switching channel; it also drives a topological reconstruction of the electronic states. The representative band structures of $Pmn2_1$ and $Abm2$ are shown in Fig.~S2~\cite{SM}. Along the high-symmetry lines of the first Brillouin zone, the low-energy band degeneracy shifts from the $\Gamma$--X direction in $Pmn2_1$ to the $\Gamma$--S direction in $Abm2$. The path-resolved band evolution near $\Gamma$ (Fig.~S3a~\cite{SM}) further shows that the anisotropy of the band splitting rotates together with the two-component order-parameter in the $(Q_a,Q_b)$ space. Polarization rotation therefore acts as a knob that controls the band gap, spin topology, and Berry geometry.
	
	Figure~\ref{fig2}(a) summarizes the evolution of the direct band gap, the projected $S_z$ spin gap, and the spin Chern number $C_s$ as functions of $\theta_Q$. Starting from $Pmn2_1$ at $\theta_Q=0^\circ$, the direct gap decreases continuously, closes near the intermediate part of the path at $\theta_Q \approx 33.5^\circ$, and then reopens toward the $Abm2$ saddle point at $\theta_Q=45^\circ$. This gap closing and reopening is accompanied by a change of the spin Chern number from $C_s=-2$ to $C_s=0$ in the gapped part of the reopened phase, demonstrating that order-parameter rotation induces a topological phase transition. The spin Chern number was obtained by diagonalizing the projected spin operator $P(\mathbf{k})S_zP(\mathbf{k})$ in the occupied subspace using the spin Wilson-loop formalism~\cite{Sheng2006PRL,Prodan2009PRB,Lange2023PRR}. When the electronic band gap is open and the projected spin spectrum remains gapped, the occupied states can be separated into spin sectors, and the Wilson-loop winding of each sector gives the corresponding spin-sector Chern number. For $Pmn2_1$, this procedure yields $C_s=-2$, consistent with previous results~\cite{Peng2024PRL,Bai2022DoubledQSHE}. Along the rotational path, the projected $S_z$ spin gap decreases continuously and collapses only near the $Abm2$ saddle, where a well-defined spin-sector decomposition is lost. Representative path-resolved gap maps and spin Wilson-loop spectra are provided in Figs.~S3b--S3d~\cite{SM}.
	
	The momentum-resolved gap maps in Figs.~\ref{fig2}(b) and \ref{fig2}(c) visualize this transition. In the $Pmn2_1$ phase, both the direct band gap and the projected $S_z$ spin gap remain finite. At the $Abm2$ saddle point, by contrast, the small-gap regions are strongly shifted and reshaped, and the projected $S_z$ spin gap closes at generic $\mathbf{k}$ points. Thus, polarization rotation does more than tune the magnitude of the gap: it reorganizes the low-energy electronic states across momentum space.

	\begin{figure}[tph!]
		\includegraphics[clip,width=1.0\linewidth]{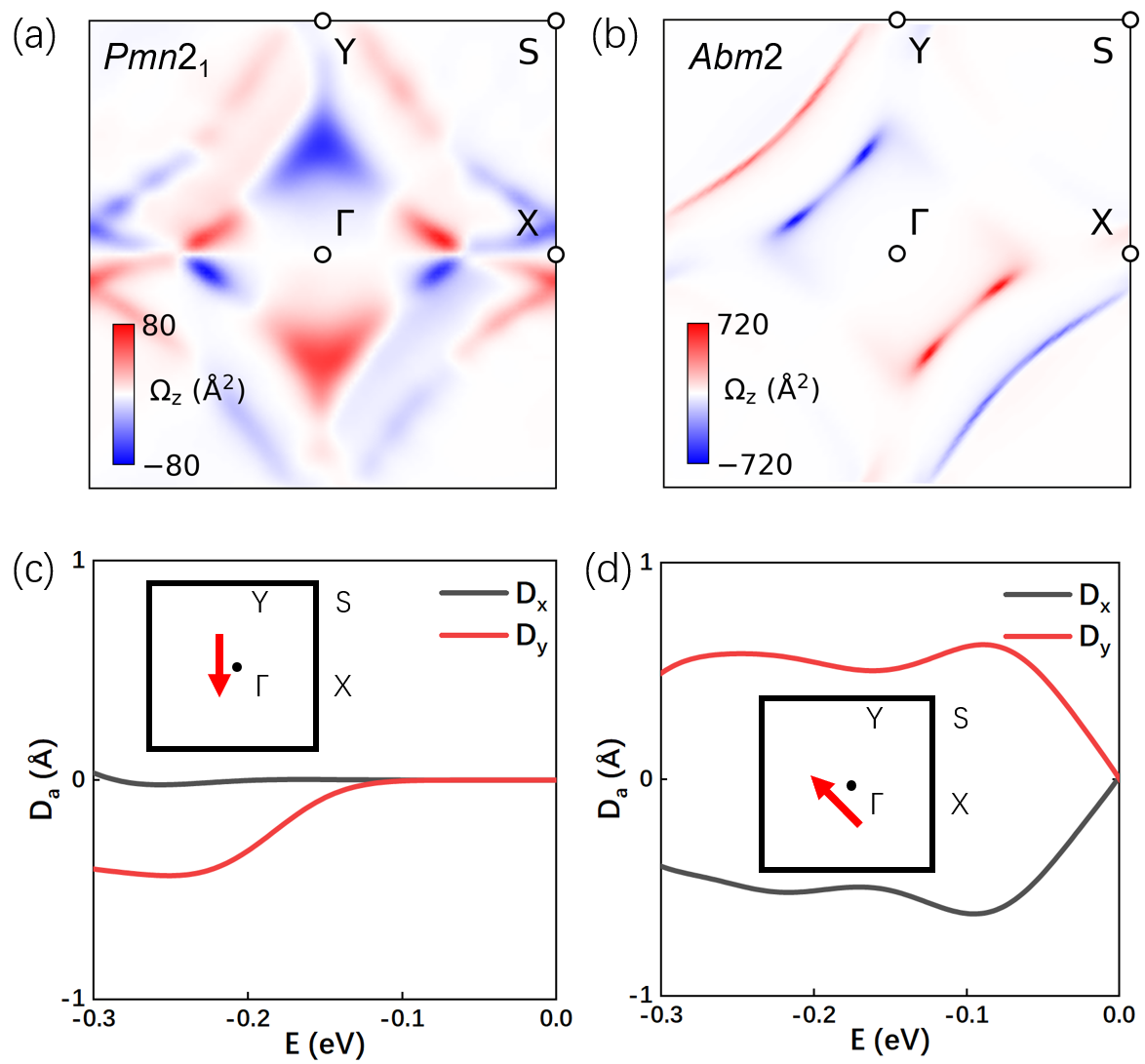}
		\caption{Berry curvature and BCD reconstruction across the rotational switching pathway. (a,b) Berry curvature distributions in the first Brillouin zone for the $Pmn2_1$ minimum and the $Abm2$ saddle point, respectively. Energy-dependent BCD components $D_x$ and $D_y$ for (c) $Pmn2_1$ and (d) $Abm2$. The red arrows indicate the direction of the BCD at $E=-0.3$ eV. The zero of energy is set to the midgap reference between the valence-band maximum and conduction-band minimum.}
		\label{fig3}
	\end{figure}

	The topological reconstruction described above leaves a direct fingerprint in the Berry geometry of the occupied and low-energy electronic states. We therefore compare the Berry curvature and BCD of the $Pmn2_1$ and $Abm2$ phases in Fig.~\ref{fig3}. In the $Pmn2_1$ phase, the Berry curvature is distributed over several low-energy regions of the Brillouin zone, with positive and negative contributions that largely compensate in the momentum-space dipole moment [Fig.~\ref{fig3}(a)]. In contrast, the $Abm2$ saddle exhibits more localized and enhanced Berry-curvature hot spots [Fig.~\ref{fig3}(b)]. This redistribution reflects the reconstruction of the spin-orbit-coupled band geometry near the spin-topological transition and provides the microscopic origin of the modified BCD. The evolution of Berry-curvature maps with $\theta_Q$ is shown in Fig.~S3e~\cite{SM}.
	
	The BCD connects this Berry-geometric reconstruction to nonlinear transport, because the leading nonlinear Hall current in time-reversal-symmetric but inversion-broken systems is proportional to the Fermi-surface BCD~\cite{Sodemann2015PRL,Ma2019Nature,Kang2019NatMater,Sinha2022NatPhys}. For a 2D system, the in-plane BCD vector $\mathbf{D}=D_x\hat{\mathbf{x}}+D_y\hat{\mathbf{y}}$ determines both the magnitude and the direction of the symmetry-allowed nonlinear Hall response. The energy-dependent BCD components in the hole-doping window from $-0.3$ to $0$~eV are shown in Figs.~\ref{fig3}(c) and \ref{fig3}(d), with $E=0$ set to the midgap energy. For the ferroelectric $Pmn2_1$ phase, we obtain a finite BCD response qualitatively consistent with previous results for doped Bi monolayer~\cite{Jin2021NanoLett}. In this energy range, the response is dominated by $D_y$. The $Abm2$ structure also preserves time-reversal symmetry while breaking inversion symmetry, and therefore supports a nonzero BCD under hole doping. However, the Berry curvature in $Abm2$ is strongly reshaped relative to $Pmn2_1$, with its principal distribution axis rotated in momentum space. For a representative hole-doped case in which the Fermi level is rigidly shifted by $-0.3$~eV, this redistribution yields a substantially enhanced BCD magnitude and a pronounced reorientation of $\mathbf{D}$, including a sign reversal of the dominant $D_y$ component and the emergence of a sizable $D_x$ component [Figs.~\ref{fig3}(c) and \ref{fig3}(d)]. Thus, the polarization-rotation pathway reconstructs not only the low-energy gap and spin topology, but also the Berry geometry governing the Fermi-surface nonlinear response. Monolayer Bi therefore provides a single-material platform in which polarization rotation can tune topology and nonlinear transport. The evolution of $D_x$ and $D_y$ with order-parameter rotation is shown in Fig.~S3f~\cite{SM}.

	\begin{figure}[tph!]
		\includegraphics[clip,width=1.0\linewidth]{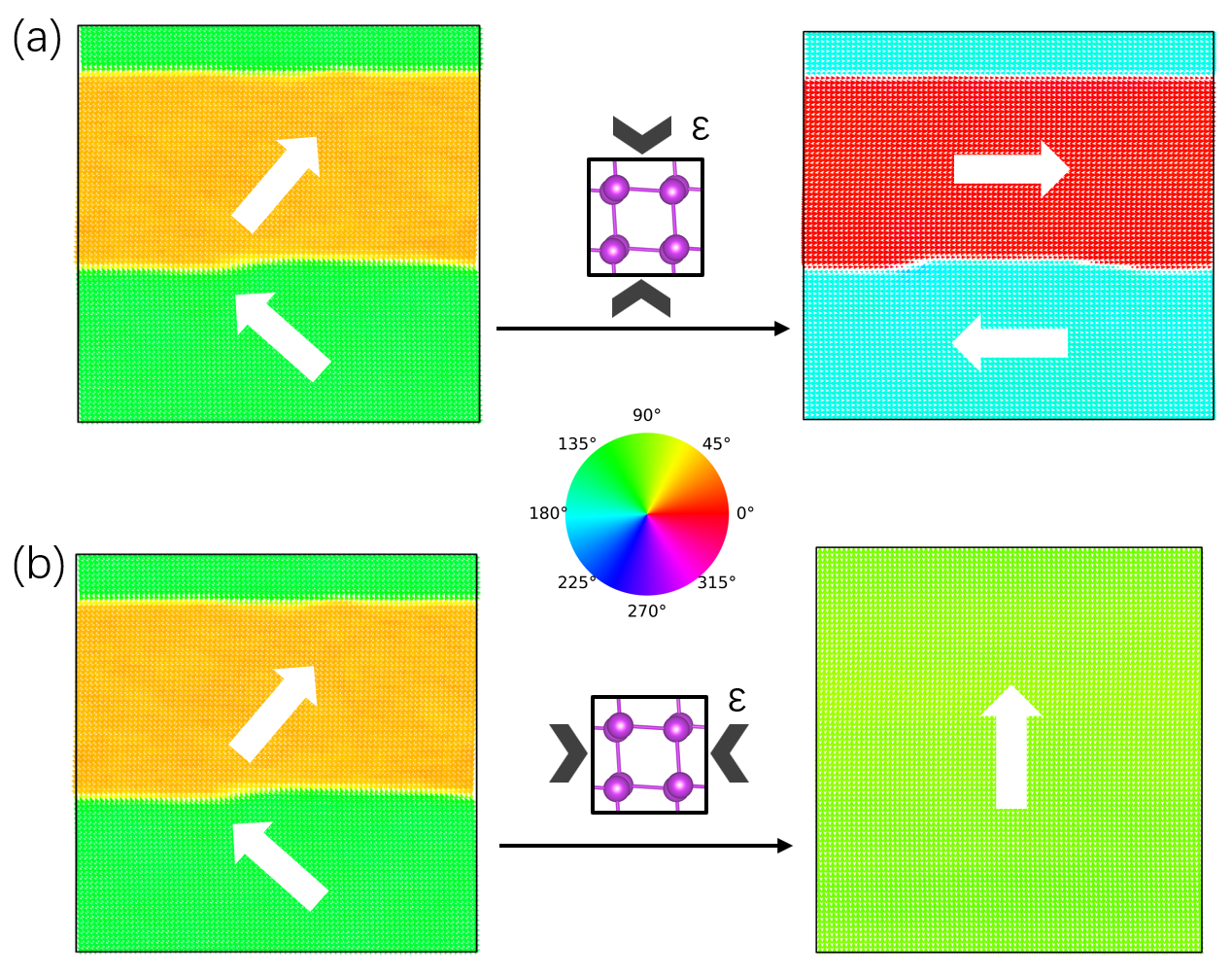}
		\caption{Strain-driven domain evolution in an optimized $Pmn2_1$ supercell containing $90^\circ$ domains under a $-5\%$ uniaxial compressive strain. The inset in each panel shows the $Pmn2_1$ unit cell whose order-parameter direction forms an angle of $45^\circ$ with the applied loading direction. (a) Compression along one direction drives the formation of $Abm2$ domains with opposite order-parameter orientations, corresponding to a $180^\circ$ domain configuration. (b) Compression along the orthogonal direction selects a single $Abm2$ orientation and produces a single-domain state.}
		\label{fig4}
	\end{figure}

	The rotational switching mechanism also points to an efficient route for external control. We therefore apply uniaxial strain along an in-plane direction $45^\circ$ away from the $Pmn2_1$ polarization direction, distinct from earlier studies in which the applied stress was parallel to the polarization direction of $Pmn2_1$~\cite{Zhong2024NatCommun,Zhong2023PRL,Wang2023PRB}, as schematically shown in Fig.~\ref{fig4}. This loading geometry is designed to couple directly to polarization rotation, rather than only to the amplitude of the polar distortion. Our MD simulations show that compressive strain in this geometry can drive a collective reorientation of the order parameter. Starting from a $Pmn2_1$ domain configuration, the polarization rotates under compression and eventually aligns with the $Abm2$ direction. This transformation is not a homogeneous unit-cell-level distortion. It proceeds through a mesoscale reorganization involving spatially extended regions with distinct order-parameter orientations, as shown by the color-coded vector maps in Fig.~\ref{fig4} and Fig.~S8~\cite{SM}. Compression along one in-plane loading direction produces $Abm2$ domains with opposite order-parameter orientations, corresponding to a $180^\circ$ domain configuration [Fig.~\ref{fig4}(a)]. By contrast, compression along the orthogonal direction selects a single $Abm2$ orientation and yields a single-domain state [Fig.~\ref{fig4}(b)]. The loading direction therefore acts as a selector for the polarization domain texture. Because the same polarization rotation also drives the spin topological transition and BCD reconstruction discussed above, uniaxial compressive strain provides a practical means to control polarization, domain configuration, electronic topology, and nonlinear Hall response in a single material. Bismuth monolayer thus realizes a mechanically programmable platform in which structural rotation, topology, and nonlinear transport are tied together by a common soft-mode degree of freedom.

	In summary, our results identify Bi monolayer as a 2D soft-mode ferroelectric whose switching coordinate is intrinsically multidimensional. The rotational channel uncovered here is more than a material-specific low-barrier path between polar states. It reveals a general mechanism in which the angular degree of freedom of a degenerate order parameter serves as a common control knob for ferroelectricity, spin topology, and Berry-geometric response. This enables topological and nonlinear Hall functionalities to be switched through polarization rotation rather than direct polarization reversal, and offers an efficient means to select among competing ferroic valleys by mechanical loading. Our additional calculations in the Supplemental Material~\cite{SM} further show that analogous $Pmn2_1$, $Pc$, and $Abm2$ branches also occur in other elemental group-VA and IV-VI monolayers. The mechanism proposed here can therefore extend beyond Bi and serve as a symmetry-based design principle for functional 2D ferroelectrics. These results establish polarization rotation as a microscopic switching mechanism in elemental group-VA monolayers and as a controllable route for manipulating topology through ferroelectric dynamics, with implications for electrically and mechanically programmable quantum and nonlinear-response devices.

	\vspace{0.5cm}
	
	\begin{acknowledgments}
		J.Z., Y.W., H.L., and J.H. acknowledge the support received from the National Natural Science Foundation of China (Grant No.\,12374024) and Fundamental Research Funds for the Central Universities (No. FRF-BRB-25-006). Y.Y. acknowledges the support received from the National Natural Science Foundation of China (Grant No.\,12304115), Fundamental Research Funds for the Central Universities (Grant No. FRF-TP-24-039A), and 2023 Fund for Fostering Young Scholars of the School of Mathematics and Physics, USTB (Grant No. FRF-BR-23-01B). Z.Y. acknowledges the support received from the National Natural Science Foundation of China (Grant No.\,12504046).
	\end{acknowledgments}
	
	\bibliographystyle{apsrev4-2}
	\bibliography{ref}
	
\end{document}